\documentclass{article}

\usepackage{arxiv}

\usepackage[utf8]{inputenc} 
\usepackage[T1]{fontenc}    
\usepackage{float}
\usepackage{hyperref}       
\usepackage{url}            
\usepackage{booktabs}       
\usepackage{amsfonts}       
\usepackage{nicefrac}       
\usepackage{microtype}      
\usepackage{lipsum}
\usepackage{graphicx}
\usepackage{amssymb}
\usepackage{amsmath}
\usepackage{bm}
\usepackage{array}
\usepackage{rotating}
\usepackage{multirow}
\usepackage{booktabs}
\usepackage{adjustbox}
\usepackage{verbatim}

\title{Trading characteristics of member firms on the Korea Exchange}

\author{
  Min-Young Lee\\
  Department of Physics\\
  Pohang University of Science and Technology\\
  Pohang 37673, Republic of Korea \\
  \texttt{minyoung@postech.ac.kr} \\
   \And
  Woo-Sung Jung\\
  Department of Physics\\
  Department of Industrial and Management Engineering\\
  Pohang University of Science and Technology\\
  Asia Pacific Center for Theoretical Physics\\
  Pohang 37673, Republic of Korea \\
  \texttt{wsjung@postech.ac.kr} \\
    \And
  Gabjin Oh \\
  College of Business\\
  Chosun University\\
  Gwangju 61452, Republic of Korea \\
  \texttt{phecogjoh@gmail.com} \\
}

\begin{document}
\maketitle

\begin{abstract}
In this paper, we study the characteristics of the member firms on the Korea Exchange. The member firms intermediate between the market participants and the exchange, and all the participants should trade stocks through members. To identify the characteristics of member firms, all member firms are categorized into three groups, such as the domestic members similar to individuals (DIMs), the domestic members similar to institutions (DSMs), and the foreign members (FRMs), in terms of the type of investor. We examine the dynamics of the member firms. The trading characteristics of members are revealed through the directionality and trend. While FRMs tend to trade one-way and move with the price change, DIMs are the opposite. In the market, DIMs and DSMs do herd and the herding moves in the opposite direction of the price change. One the other hand, FRMs do herd in the direction of the price change. The network analysis supports that the members are clustered into three groups similar to DIMs, DSMs, and FRMs. Finally, random matrix theory and a cross-sectional regression show that the inventory variation of members possesses significant information about stock prices and that member herding helps to price the stocks.
\end{abstract}


\section{INTRODUCTION}

The stock market is a complex system in which a large number of investors with heterogeneous strategies participate. The trading activities of many investors make stock prices fluctuate. Analyzing the trading strategies of participants helps us to understand the movement of stock prices. The Korea Exchange (KRX) has the 14th largest market capitalization in the world, and the market is a representative emerging market with different characteristics from a developed market. Individual investors actively participate in the market. The smaller the market capitalization is, the greater the proportion of individuals. In the US market, which is a developed market, the individual investor owned approximately 90\% of the stock in 1950. The individual investor owned approximately 30\% of the stock in 2009, and only approximately 2\% of the trading volume in the NYSE is traded by individual investors\cite{Evans2009}. On the other hand, 60.3\% of the transaction amount in the KRX from 2007 to 2017 was traded by individual investors. Even individuals account for more than 90\% of the trading of stocks with small market capitalizations.
We analyze the characteristics of the member firms in the KRX. Member firms are the companies that trade on behalf of their clients or trade their accounts. The clients are categorized as individuals, institutions, and foreigners. Member firms are only entitled to trade on the exchange; therefore, all investors have to go through their member firms when trading stocks. Because heterogeneous investors trade through member companies, identifying the characteristics of member companies is a difficult problem. In addition, the KRX rarely discloses information about member companies. Through the daily transaction of member firms, however, it is possible to get a lot of information about the member firms indirectly. The characteristics of member firms are determined by the type of clients, such as individuals, institutions, and foreigners, that trade through their member firm. Few studies have dealt with the characteristics of the identified member firms. To the best of our knowledge, this study is the first one analyzing the trading behavior of the identified member firms on the Korean stock market. 
The characteristics of individuals, institutions, and foreigners is a subject that many researchers have studied. Gabaix et al.\cite{Gabaix2006} showed the excess volatility from large institutions in illiquid markets. Bohl et al.\cite{Bohl2006} found no evidence that institutions destabilize the market. Barber et al.\cite{Barber2000,Barber2009,Barber2013} investigated the investment performance of individuals. Foucault et al.\cite{Foucault2011} showed the effect of individual trading on volatility. Adaoglu et al.\cite{Adaoglu2013} showed the cause-and-effect between stock returns and foreign investor flows. Bae et al.\cite{Bae2011} showed the superior performance of foreign investors. 

Much research has been conducted on the KRX, which has different characteristics from a developed market. Park and Kim studied the performance of individual traders \cite{Park2014} on the KRX. In addition to studies about individuals, studies have also been conducted on institutions and foreigners. There have been studies focused on the trading of foreign investors on the KRX based on herding and positive feedback \cite{Choe1999,Jeon2010}. However, as far as we know, there have been no studies of member firms on the KRX. The ambiguity about member transactions is resolved indirectly. Thus, we have identified the member's trading characteristics using the type of clients.
We analyzed the transaction characteristics of member firms. First, we show the similarity between member firms and investor types such as individuals, institutions, and foreigners. All member firms are categorized into three groups, such as domestic members similar to individuals (DIMs), domestic members similar to institutions (DSMs), and foreign members (FRMs) in terms of the type of investor. We introduce the measures of the directionality and trend to characterize the trading behavior of members. From the measures, it can be seen that the FRMs trade in the same direction as the changes in stock prices and conduct their intraday trading in one direction.
Herding is investment that mimics other investors and can result from rational or irrational transactions. Herding does not necessarily reverse an unusual return. One of the interesting characteristics of member firms on the KRX is herding in the opposite direction. In general, investor herding makes a price change in the same direction. The buy herding of investors, for example, makes a return turn positive. In the KRX, however, the herding of member firms makes a price change in the opposite direction. While the herding of FRMs moves in the direction of the price change. We also construct the network based on the correlation of the inventory variation of members. The community detection algorithm shows that the network is divided into groups that are almost similar to the previous ones regarding the characteristics of members: DIMs, DSMs, and FRMs.
The final step is to find a link between the trading of members and stock prices. Random matrix theory shows that the inventory variation possesses information about stock prices. Cross-sectional regression demonstrates that the herding of members complements the market factor to explain stock prices. The inventory variation of members possesses significant information on stock prices.
Our research contributes four points to this research field. First, we find that member firms do herd in the opposite direction on the KRX. Second, there have been few studies on the characteristics of member firms on the Korea Exchange. We have identified the characteristics of member firms compared with the investor types and used measures such as the directionality and trend. Third, from the community structure of the member network, we show how herding takes place at the level of each member. Finally, random matrix theory and a cross-sectional regression show that the inventory variation and herding possess a lot of information about the price dynamics and help price stocks.

The remainder of this paper is organized as follows. In section 2, we explain the KRX dataset. In section 3, we dissect the inventory variation of member firms at the level of investor types and analyze the characteristics of the transactions of each member. In section 4, the herding of members on the KRX is introduced. In section 5, the community structure of the member network is investigated. In section 6, we study the relationship and correlation between the inventory and stock returns using random matrix theory and a cross-sectional regression.

\section{Data}

The dataset we use consists of the daily prices and inventory variations on the Korea Exchange (KRX). The inventory variation is the buy and sell transaction amount, and is the product of the transaction volume and price. The period of the dataset is from 1 January 2007 to 31 December 2017 (11 years, 2722 trading days) and the time resolution is one day. The 1210 firms that are continuously traded from 2007 to 2017 are included. The firms are categorized into ten deciles according to their market capitalization. Unlike developed markets, individuals are very active on the KRX. Table 1 shows the transaction amounts and the ratios of the investors by market capitalization. The transaction amount represents the number of traded shares multiplied by the price. Approximately 43.11\% of the transaction amount is traded by individual investors in the first decile of market capitalization. The lower the market cap is, the greater the ratio of individual trading. In the tenth decile group, the ratio increases to 96.54\%. We use the inventory variation of 62 member firms and the inventory variation of 3 investor types (individuals, institutions, and foreigners). The list of members firms is shown in Table 2. Nos. 1 - 41 are the domestic members and Nos. 42 - 62 are the foreign members. KRX categorizes investors into individuals, institutions, and foreigners and provides daily trading information. Because only the member firms are entitled to trade on the exchange, all investors have to make transactions through member firms. There are 41 domestic members and 21 foreign members on the KRX. The members are categorized into domestic members and foreign members according to whether the headquarters of each member is in Korea or another country. Since many investors make transactions through member firms, it is hard to consider them as independent investors. However, by analyzing the investor types that trade through the member firms, the characteristics of the member firms can be known. Domestic member firms can be explained as having a combination of individuals and institutions. Foreign member firms can be described as foreign investors. In 2003, 99.8\% of the foreign investors were foreign institutional investors\cite{Jeon2010}.

\begin{table}[tbp]
\caption{Total transaction amount ($10^{15}$ KRW) of each investor type from Jan. 2007 to Dec. 2017. The numbers in parentheses indicate the proportion of the transaction amount of each investor. The 1210 stocks are divided into decile groups based on market capitalization. The first decile is the firms with the 121 largest market capitalizations.}
\begin{adjustbox}{width=15cm,angle=0}
\begin{tabular}{c|cccccccccc|c}
\hline
\multirow{2}{*}{Investor type} & \multicolumn{10}{c|}{Market capitalization decile}                                               &         \\ \cline{2-12} 
                               & 1       & 2       & 3      & 4       & 5       & 6       & 7       & 8       & 9       & 10      & Total   \\ \hline
Individual                     & 81.17   & 21.1    & 15.07  & 12.43   & 13.03   & 10      & 9.64    & 8.07    & 7.19    & 5.45    & 183.15  \\
(\%)                           & (43.11) & (77.15) & (85.9) & (88.98) & (91.17) & (92.59) & (94.85) & (96.44) & (96.33) & (96.54) & (60.27) \\ \hline
Foreigner                      & 55.98   & 2.66    & 1.15   & 0.73    & 0.64    & 0.46    & 0.31    & 0.21    & 0.18    & 0.13    & 62.45   \\
(\%)                           & (29.73) & (9.72)  & (6.58) & (5.21)  & (4.48)  & (4.22)  & (3.04)  & (2.46)  & (2.44)  & (2.34)  & (20.55) \\ \hline
Institution                    & 51.12   & 3.59    & 1.32   & 0.81    & 0.62    & 0.34    & 0.22    & 0.09    & 0.09    & 0.06    & 58.26   \\
(\%)                           & (27.15) & (13.14) & (7.52) & (5.81)  & (4.35)  & (3.18)  & (2.12)  & (1.1)   & (1.23)  & (1.13)  & (19.17) \\ \hline
Total                          & 188.27  & 27.35   & 17.54  & 13.97   & 14.29   & 10.8    & 10.17   & 8.37    & 7.46    & 5.64    & 303.86  \\ \hline
\end{tabular}
\end{adjustbox}
\end{table}

\begin{sidewaystable}
\centering
\caption{Statistics for member firms on the KRX. The period is the time in which the member traded, and $N_d$ is the number of trading days. $N_s$ is the number of stocks that a member trades in the total period. Vol is the total transaction amount from January 2007 to December 2017 and the unit is .}
\small
\begin{adjustbox}{width=20cm,center}
\begin{tabular}{llllll|llllll}
\hline
No & Name           & Period ($N_d$)     & $N_s$ & Vol     & Vol/$N_{ds}$ & No & Name               & Period ($N_d$)     & $N_s$ & Vol     & Vol/$N_{ds}$ \\ \hline
1  & Kyobo          & 0701-1712 (2722) & 1210 & 368.57  & 112       & 32 & CAPE               & 0807-1712 (2336) & 1210 & 82.65   & 29        \\
2  & Shinhan Invest & 0701-1712 (2722) & 1210 & 1554.91 & 472       & 33 & BNK                & 0912-1712 (1987) & 1210 & 41.64   & 17        \\
3  & Korea Invest   & 0701-1712 (2722) & 1210 & 1789.34 & 543       & 34 & IM                 & 0701-1509 (2154) & 1207 & 47.1    & 18        \\
4  & Daishin        & 0701-1712 (2722) & 1210 & 1115.13 & 339       & 35 & NH (Nonghyup)      & 0701-1505 (2071) & 1210 & 221.41  & 88        \\
5  & Mirae Daewoo   & 0701-1712 (2722) & 1210 & 1993.64 & 605       & 36 & KB Invest          & 0701-1705 (2567) & 1210 & 428.53  & 138       \\
6  & Shinyoung      & 0701-1712 (2722) & 1210 & 187.56  & 57        & 37 & Mirae Asset        & 0701-1612 (2479) & 1210 & 2002.43 & 668       \\
7  & Eugene         & 0701-1712 (2722) & 1210 & 418.8   & 127       & 38 & Hanwha Invest      & 0701-1208 (1412) & 1210 & 75.05   & 44        \\
8  & Hanyang        & 0701-1712 (2722) & 1210 & 106.51  & 32        & 39 & BNG                & 0701-1411 (1913) & 1198 & 24.04   & 10        \\
9  & Meritz         & 0701-1712 (2722) & 1210 & 310.94  & 94        & 40 & Apple              & 0808-1311 (1290) & 1164 & 12.94   & 9         \\
10 & NH Invest      & 0701-1712 (2722) & 1210 & 1808.78 & 549       & 41 & Hanmag             & 0902-1401 (1219) & 1170 & 22.63   & 16        \\
11 & Bookook        & 0701-1712 (2722) & 1209 & 88.63   & 27        & 42 & JP Morgan *          & 0701-1712 (2722) & 1193 & 353.63  & 109       \\
12 & KB             & 0701-1712 (2722) & 1210 & 1350.58 & 410       & 43 & Macquarie *          & 0701-1712 (2722) & 919  & 261.92  & 105       \\
13 & Hanwha         & 0701-1712 (2722) & 1210 & 533.85  & 162       & 44 & Morgan Stanley *     & 0701-1712 (2722) & 1207 & 794.98  & 242       \\
14 & Hyundai Motor  & 0701-1712 (2722) & 1210 & 202.37  & 61        & 45 & Citigroup *          & 0701-1712 (2722) & 981  & 434.7   & 163       \\
15 & Yuhwa          & 0701-1712 (2722) & 1210 & 40.29   & 12        & 46 & HSBC *               & 0701-1712 (2722) & 900  & 111.2   & 45        \\
16 & Yuanta         & 0701-1712 (2722) & 1210 & 1206.15 & 366       & 47 & CLSA *               & 0701-1712 (2722) & 939  & 333.71  & 131       \\
17 & SK             & 0701-1712 (2722) & 1210 & 525.92  & 160       & 48 & Credit Suisse *      & 0701-1712 (2722) & 1208 & 833.59  & 254       \\
18 & Golden Bridge  & 0701-1712 (2722) & 1210 & 107.24  & 33        & 49 & UBS *                & 0701-1712 (2722) & 1195 & 512.25  & 157       \\
19 & Samsung        & 0701-1712 (2722) & 1210 & 1907.22 & 579       & 50 & Merrill Lynch *      & 0701-1712 (2722) & 1205 & 636.5   & 194       \\
20 & DB Financial   & 0701-1712 (2722) & 1210 & 422.38  & 128       & 51 & Goldman Sachs *      & 0701-1712 (2722) & 1203 & 491.1   & 150       \\
21 & HI Invest      & 0701-1712 (2722) & 1210 & 321.7   & 98        & 52 & Societe Generale *   & 0701-1712 (2722) & 947  & 107.49  & 42        \\
22 & Kiwoom         & 0701-1712 (2722) & 1210 & 3969.51 & 1205      & 53 & Nomura *             & 0701-1712 (2722) & 1004 & 154.34  & 56        \\
23 & Leading Invest & 0701-1712 (2722) & 1206 & 56.88   & 17        & 54 & Deutsche *           & 0701-1712 (2722) & 1193 & 382.49  & 118       \\
24 & Hana Financial & 0701-1712 (2722) & 1210 & 917.87  & 279       & 55 & Daiwa *              & 0701-1712 (2722) & 876  & 70.98   & 30        \\
25 & eBEST          & 0701-1712 (2722) & 1210 & 531.28  & 161       & 56 & BNP Paribas *        & 0701-1712 (2722) & 855  & 106.72  & 46        \\
26 & Korea Asset    & 0701-1712 (2722) & 1189 & 37.05   & 11        & 57 & Standard Chartered * & 0808-1501 (1292) & 390  & 10      & 20        \\
27 & Heungkuk       & 0701-1712 (2722) & 1117 & 44.98   & 15        & 58 & CIMB *               & 1302-1712 (1175) & 683  & 12.79   & 16        \\
28 & IBK            & 0807-1712 (2336) & 1210 & 128.6   & 45        & 59 & RBS *                & 0701-1203 (1301) & 658  & 35.32   & 41        \\
29 & Baro Invest    & 0807-1712 (2336) & 1129 & 32.57   & 12        & 60 & Newedge *            & 0701-1412 (844)  & 120  & 1.69    & 17        \\
30 & Taurus Invest  & 0807-1712 (2336) & 1204 & 49.42   & 18        & 61 & Barclays *           & 1108-1602 (1057) & 724  & 30.31   & 40        \\
31 & KTB            & 0807-1712 (2334) & 1210 & 149.51  & 53        & 62 & ING *                & 0906-1303 (214)  & 58   & 0.03    & 2         \\ \hline
\end{tabular}
\end{adjustbox}
\end{sidewaystable}

\section{Investor type of members}

A description of the types of investors that make up the Korean stock market will help to understand the member firms. The market has three major investor categories: individuals, institutions and foreigners. In this context, the institutions represent the domestic institutions. To understand the characteristics of member firms, it is necessary to know the three investor categories. The relation between the three investor types is analyzed using the Pearson correlation coefficient and the partial correlation coefficient as follows. Since the three investors influence each other, the partial correlation coefficient was introduced to determine the correlation between two variables and to control the other variable. When comparing individuals and foreigners, for example, institutional transactions affect the transaction of individuals and foreigners. For that reason, the partial correlation is used to exclude the effects of the institutional transactions.

\begin{align}
{\rho}_{XY} = \frac{N\Sigma^{N}_{i=1}X_{i}Y_{i}-\Sigma^{N}_{i=1}X_{i}\Sigma^{N}_{i=1}Y_{i}}{\sqrt{N\Sigma^{N}_{i=1}X^2_{i}-\left ( \Sigma^{N}_{i=1}X_{i} \right )^2}\sqrt{N\Sigma^{N}_{i=1}Y^2_{i}-\left ( \Sigma^{N}_{i=1}Y_{i} \right )^2}}
\end{align}

\begin{align}
&\textbf{w}^*_X = \underset{\textbf{w}}{\mathrm{argmin}}\left \{ \sum_{i=1}^{N}(x_i-\left \langle \textbf{w},\textbf{z}_i \right \rangle)^2 \right \} \\
&\textbf{w}^*_Y = \underset{\textbf{w}}{\mathrm{argmin}}\left \{ \sum_{i=1}^{N}(y_i-\left \langle \textbf{w},\textbf{z}_i \right \rangle)^2 \right \} \nonumber \\
&e_{X,i} = x_i - \left \langle \textbf{w}^*_X,\textbf{z}_i \right \rangle \\
&e_{Y,i} = y_i - \left \langle \textbf{w}^*_Y,\textbf{z}_i \right \rangle \nonumber \\
&\rho_{XY\cdot \textbf{Z}} = \rho_{e_X,e_Y}
= \frac{N\Sigma^{N}_{i=1}e_{X,i}e_{Y,i}}{\sqrt{N\Sigma^{N}_{i=1}e^2_{X,i}}\sqrt{N\Sigma^{N}_{i=1}e^2_{Y,i}}}
\end{align}

$\rho_{XY}$ is the Pearson correlation between variables X and Y and $\rho_{XY\cdot \textbf{Z}}$ is the partial correlation between $X$ and $Y$ controlling the other variable $Z$. $N$ is the number of observations and $\textbf{w}^{*}$ is the regression coefficient vectors. $\left \langle \textbf{w}^*_X,z_i \right \rangle$ is the scalar product between $a$ and $b$. $e_{X,i}$ and $e_{Y,i}$ are the residuals. 

The Pearson correlation coefficient shows that the individuals have a negative correlation with institutions and foreigners; however, there is an insignificant correlation between the institutions and foreigners. The partial correlation coefficient that controls the variable of individuals reveals that institutions and foreigners also have a negative correlation. Fig. 1 (a)-(c) shows the scatter plot of Samsung Electronics (005930) as an example. Fig. 1 (d)-(f) show the correlation and partial correlation coefficients for the inventory variations of the three investor types for all 1210 stocks.

Next, we investigate the relationship between inventory variations and price returns. For the large market caps, the inventory variations of the institutions and foreigners have a positive correlation with the price return. The inventory variation of individuals, however, has a negative correlation with the price return. For the stocks with small market caps, the correlation between the price return and inventory variation converges to zero. Fig. 2 shows the correlation between the inventory variations of the three types and the price returns by market capitalization decile.

\begin{figure}[h!]
\includegraphics[scale=0.4]{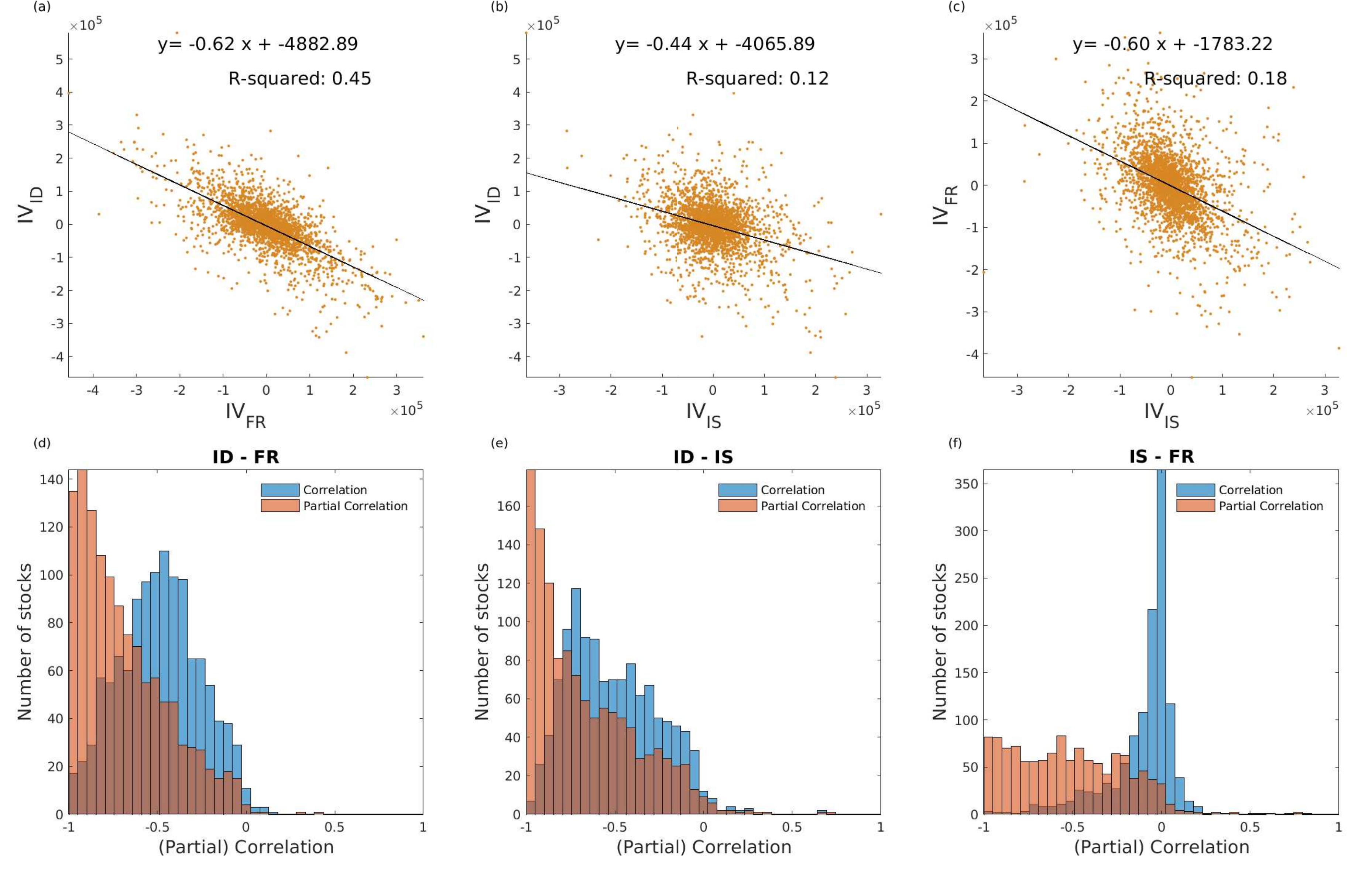}
\caption{(a)-(c) Scatter plot for the inventory variations of two variables for Samsung Electronics (005930). (a) Individuals(ID) and Foreigners(FR), (b) Individuals(ID) and Institutions(IS), (c) Foreigners(FR) and Institutions(IS). (d)-(f) Distributions of the Pearson correlation and partial correlation coefficients for the inventory variations of 1210 stocks. (d) ID and FR. (Control: IS) (e) ID and IS. (Control: FR) (f) IS and FR. (Control: ID)}
\end{figure}

\begin{figure}[h!]
\includegraphics[scale=0.7]{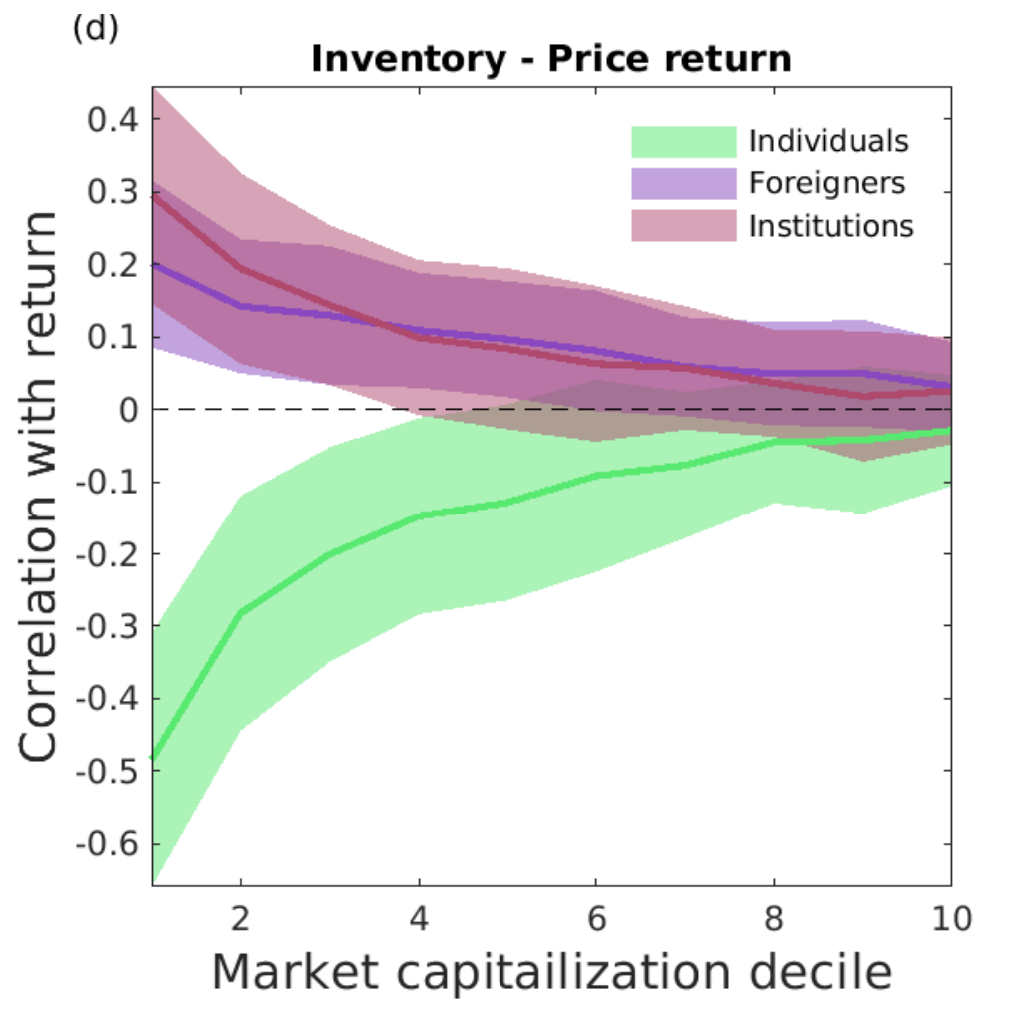}
\caption{Correlation between the inventory variations and price returns for 1210 stocks. The shaded area represents the standard deviation of the decile.}
\end{figure}

The classification of a member firm is determined by the ratio of the investor types. Individuals and institutions trade through the same domestic member firms. On the other hand, foreigners usually trade through foreign member firms. Unfortunately, the KRX provides no information about the portion of individuals, foreigners, and institutions that trade through each member firm. We indirectly identify the portion of the trader types in Fig. 3 using the correlation between the inventory variation of members and the inventory variations of individuals, foreigners, and institutions. As shown in Fig. 2, the trading behaviors of the three investor types are clearly distinguished for the first decile of market capitalization; thus, we use the first decile to investigate the member firms. For example, in Fig. 3 (a), Kiwoom and have very high proportions of individual investors, whereas KTB and HI Invest have small portions of individual traders and relatively high portions of institutions. In general, the members with higher proportions of individuals are the larger firms.

Based on the relative proportion of individuals and institutions, we classify the domestic member firms into three categories. Domestic members with individuals dominant (DIMs) are on the right bottom in Fig. 3 (a): Shinhan Invest, Korea Invest, Daishin, Mirae Daewoo, NH Invest, KB, Yuanta, Samsung, Kiwoom, eBEST, Mirae Asset, and Hanwha Invest. Domestic members with institutions dominant (DSMs) are on the left top in Fig. 3 (a): Shinyoung, Hanyang, Meritz, Bookook, Yuhwa, Golden Bridge, HI Invest, Baro Invest, Taurus Invest, KTB, CAPE, BNK, IM, and KB Invest. The Foreign member firms (FRMs) are in Fig. 3 (b): JP Morgan, Macquarie, Morgan Stanley, Citigroup, HSBC, CLSA, Credit Suisse, UBS, Merrill Lynch, Goldman Sachs, Societe Generale, Nomura, Deutsche, Daiwa, BNP Paribas, Standard Chartered, CIMB, RBS, Newedge, Barclays, and ING.

\begin{figure}[h!]
\hspace*{-3cm}
\includegraphics[scale=0.37]{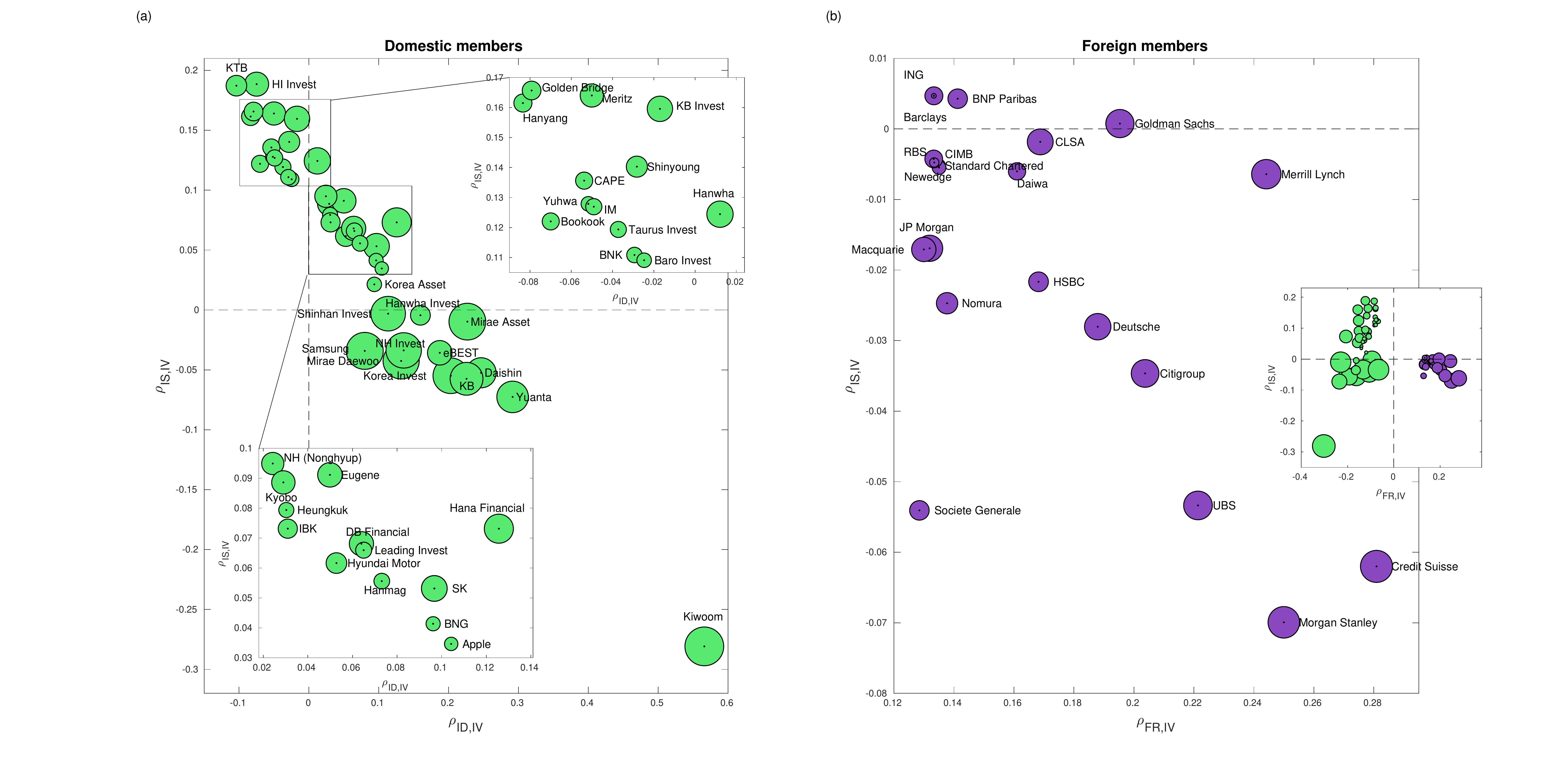}
\caption{(a) Mean of the correlation between the inventory variation of domestic members (green circle) and the inventory variation of individuals and institutions. The inset figures (right upper and left lower) show the enlarged views. (a) DIMs are on the right bottom, and DSMs are on the left top. (b) Mean of the correlation between the inventory variation of foreign members (purple circle) and the inventory variation of individuals and foreigners. The inset on (b) shows the domestic members (green circle) and foreign members (purple circle) at once. The correlations are averaged over the largest market capitalization decile. The radius is proportional to the square root of the transaction amount.}
\end{figure}

To understand the characteristics of member firms, we analyze the directionality and the trading trend of each member. We also investigate the directionality and the trading trend of the individuals, institutions, and foreigners that trade through member firms and compare them with member firms. We define the directionality and trend to measure the characteristics of the trading behavior as follows. Tumminello et al. categorized the states into the primarily buying, primarily selling and buying and selling \cite{Tumminello2012}. Similarly, we define the directionality ($D$), which indicates how much a daily transaction consistently appears. A large directionality means that most of the daily transaction amount has the same direction. The trend ($T$) represents the relative change in the inventory versus the price movement.

\begin{eqnarray}
    D_j = \frac{1}{N} \sum _{i}^{N} P(|\frac{B_{j,i}-S_{j,i}}{B_{j,i}+S_{j,i}}|\geq\theta)\\
    T_j = \frac{1}{N} \sum _{i}^{N} \frac{E[(x_{j,i}-\mu _{x_{j,i}})(r_i-\mu _{r_i})]}{\sigma _{x_{j,i}}\sigma _{r_i}}.
\end{eqnarray}

$where\; x_{j,i}=B_{j,i}-S_{j,i},\; \theta=0.2,\; and\ \; N=121$, which is the number of stocks in one decile. $B_{j,i} (S_{j,i})$ is the time series of the transaction amount of a member firm or investor $j$ to buy(sell) stock $i$. $r_i$ is the return of stock $i$ and $\mu_y$ is the mean of the variable $y$. The average length of time series $B_{j,i} (S_{j,i})$ is 247 for each year. $j$ is one of the member firms or one of the individuals, foreigners or institutions. $T_j>0$ means that member firm or investor $j$ trades stock in the trending way. Tumminello et al. used the states of a primarily buying state, a primarily selling state and a buying and selling state according to the relationship between $\frac{B-S}{B+S}$ and $\theta$ \cite{Tumminello2012}. Similarly, we define the directionality according to the relationship between $\frac{B-S}{B+S}$ and $\theta$.

Fig. 4 (a) shows that the directionalities ($D$) and trends ($T$) of individuals, foreigners, and institutions have different characteristics. For the stocks with large market capitalizations, institutions and foreigners trade in the trending way in which an inventory change moves in the same direction as a price change. Meanwhile, individuals trade in the reverse way in which an inventory change moves in the opposite direction of the price change. However, the difference in the trend gets smaller for stocks with small market capitalizations. When the market capitalization changes, the directionality and trend move in opposite directions. The directionality of individuals, foreigners, and institutions is around $0.5$ for larger market capitalizations. However, the directionality of foreigners and institutions increases for those with small market capitalizations. Conversely, Fig. 4 (b) shows the directionality and trading trend of each member for the first decile’s 121 stocks. The members which trade less than a tenth of the average are excluded. Each point in Fig. 4 (b) can be roughly classified into three types. The red circles are the DIMs, as shown in Fig. 3 (a). The green circles are the DSMs. The purple circles are the FRMs. The DIMs tend to trade against the stock price movement and do not trade in one direction. On the other hand, the FRMs tend to trade in the same direction as the price movement and trade in one direction. Fig. 4 (c) shows the results of the tenth decile. Unlike the first decile, there is almost no difference between the members.

\begin{figure*}[h!]
\hspace*{-1.3cm}
\includegraphics[scale=0.3]{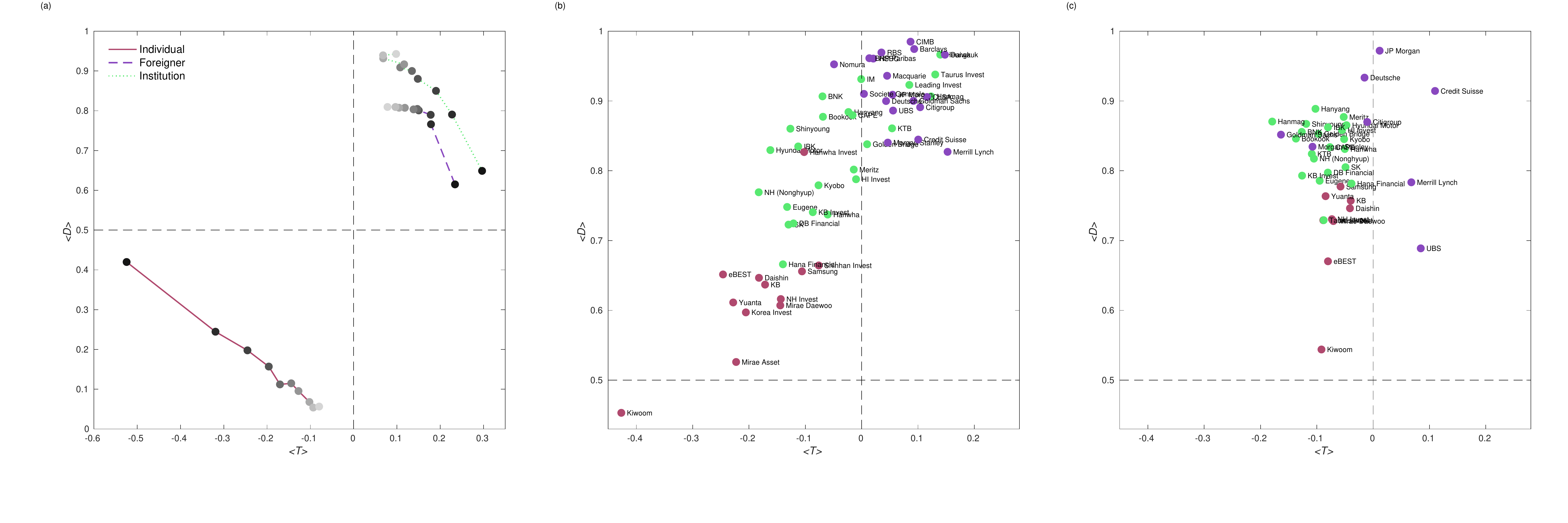}
\caption{(a) Directionalities ($D$) and trends ($T$) of individual (red), foreigner (purple), and institution (green) investors. $\theta = 0.2$. The darker color refers to the larger market capitalization decile. (b) Directionalities ($D$) and trends ($T$) of the members for the first decile stocks. Red points represent DIMs, the green points represent DSMs and the purple points represent FRMs. (c) Directionalities ($D$) and trends ($T$) of the members for the tenth decile stocks.}
\end{figure*}

\section{Herding of members}

Investors who practice herding mimic the investment decisions of others or exploit similar information as others. The behavior of herding comes from a rational or an irrational decision. In financial markets, herding can impact stock prices~\cite{Nofsinger1999,Chang2000,Sias2004,Cajueiro2009,Shyu2010,Park2011,Dasgupta2011,Blasco2012,Merli2013,Kremer2013,Spyrou2013,Cai2019}. Nofsinger et al.~\cite{Nofsinger1999} showed the impacts of herding and the positive feedback of institutional investors. Sias~\cite{Sias2004} verified that institutional investors follow each other and follow their past trading. Kremer et al.~\cite{Kremer2013} verified the existence of herding on a daily time scale. The strength of herding depends on the volatility. Choe~\cite{Choe1999} et al. analyzed the Korean stock market and found herding and the positive feedback trading of foreigners.

When investors do buy(sell) during herding, in general, the price goes up(down). The herding of member firms on the KRX, however, shows the opposite phenomena. To analyze the herding of member firms, we introduce the herding indicator $h$, the herding indicator with sign $H$ and the direction of herding $DH$ as follows, which are modified from the definitions of Zhou et al.\cite{Zhou2012} from which it is possible to consider the direction of herding:

\begin{eqnarray}
    h_{i,d} = P\Bigg(f(k,n,p) \leq 0.05\Bigg)\\
    f(k,n,p)=\binom{n}{k}p^k(1-p)^{n-k} \\
    H_{i,d} = h_{i,d} \times sign(N_{B,i,d} - N_{S,i,d}) \\
   DH_i = \frac{E[(H_{i}-\mu _{H_{i}})(r_i-\mu _{r_i})]}{\sigma _{H_{i}}\sigma _{r_i}}
\end{eqnarray}
where $f(k,n,p)$ is a probability mass function for a binomial distribution $f(k,n,p)=\binom{n}{k}p^k(1-p)^{n-k}$ and $k=N_{B,i,d}$, $n=N_{B,i,d}+N_{S,i,d}$, $p=1/2$. $N_{B,i,d} (N_{S,i,d})$ is the number of member firms that buy(sell) stock $i$ on day $d$. $r_i$ is the price return for stock $i$. $h_{i,d}$ is the herding indicator compared to the binomial null hypothesis. $h_{i,d}=1$ means the member firms herd the $i$th stock on day $d$. $H_{i,d}$ is the herding indicator with the sign. If the number of members buying the $i$th stock is equal to the numbers selling the stock on day $d$, $H_{i,d}=0$. $H_i$ is the time series of $H_{i,d}$. Because we divide the whole time series into 11 years sub time series, the average length of each $H_{i}$ is 247. $H_{i,d} =1$ refers to buy herding and $H_{i,d} =-1$ refers to sell herding. $\mu_{H_i} (\sigma_{H_i})$ is the mean (standard deviation) of $H_i$. $\mu_{r_i} (\sigma_{r_i})$ is the mean (standard deviation) of $r_i$, where $r_i$ is the price return of stock $i$. We define $DH(i)>0$ as herding and $DH(i)<0$ as herding in the opposite direction. Fig. 5 (a) shows that all member firms herd. FRMs also have weak herding, whereas the herding of DIM is relatively stronger than that of the others. Fig. 5 (b) shows that the buy(sell) herding of all member firms makes the price goes down(up). FRMs herd in the direction of the price change; otherwise, domestic members herd in the opposite direction. The main reason for the herding in the opposite direction is due to the domestic members.

\begin{figure}[h!]
\includegraphics[scale=.55]{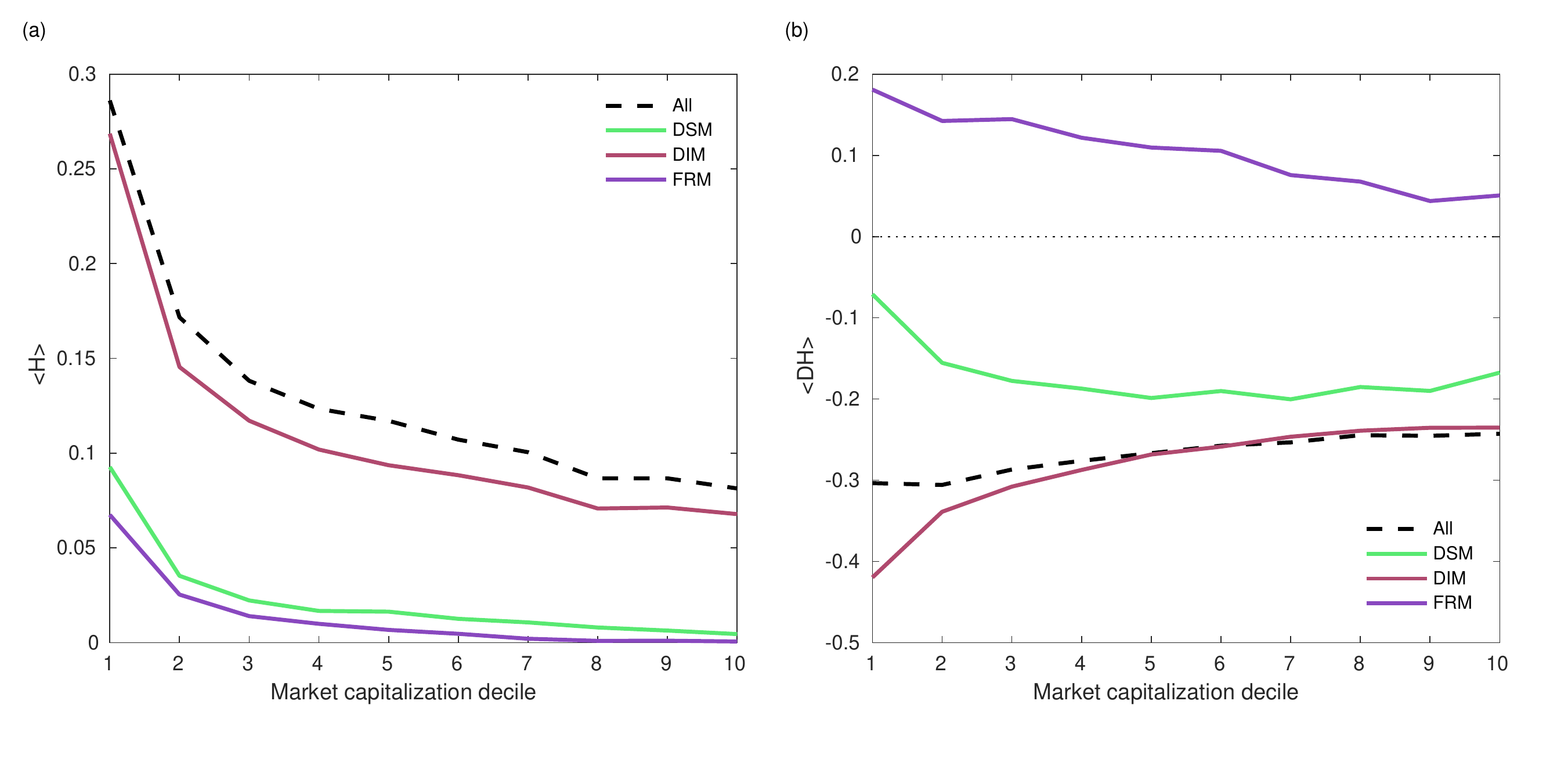}
\caption{(a) Probability of herding compared to the binomial null hypothesis. (b) Direction of herding of member firms. (a)-(b) All member firms (dotted black line), DIMs (red line), DSMs (green line), FRMs (purple line). All values are averaged over 121 stocks in the market capitalization decile and 11 years.}
\end{figure}

\begin{figure}[h!]
\hspace*{-3cm}
\includegraphics[scale=0.57]{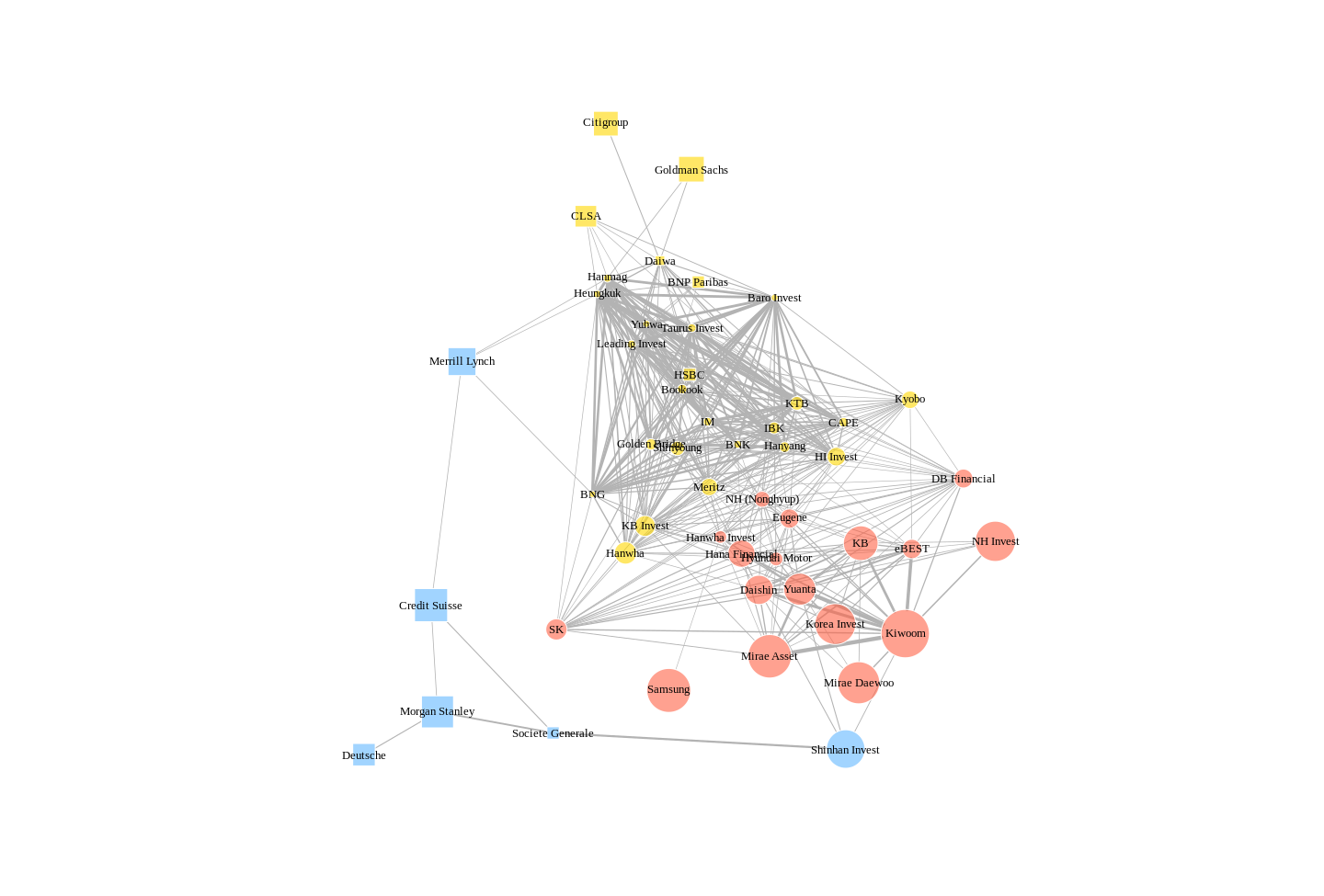}
\caption{Network and community structure of members for the first decile group. The domestic members are represented by circles and the foreign members are represented by squares. The size of a node is proportional to the square root of the transaction amount. The color of a node refers the community groups using the Infomap community detection method.}
\end{figure}

\section{Community structure of members}
There have been several studies about the clustering of investors \cite{Schweitzer2009,Jiang2010,Tumminello2012,Musciotto2016,Xie2016}. Musciotto et al. showed the hierarchical structure and cluster of investors in the Finnish market \cite{Musciotto2016}. Schweitzer et al. analyzed the financial network of financial institutions. We construct a network based on the inventory variation correlation. Apple, Hanmag, Standard Chartered, CIMB, RBS, Newedge, Barclays, and ING are excluded from the network because these members have been on the Korean market for a short time and their transaction volumes are small, which distort the network. The correlation matrix is made using the daily inventory variation of member firms. The nodes of the network correspond to each member. The strength of an edge is the mean of the correlation coefficient of the inventory variation. Domestic members are represented by circles and foreign members are represented by squares, as shown in Fig. 6. The size of a node is proportional to the square root of the trading volume of each member. When the members are separated and integrated, these are regarded as other member firms. Members who do not exist at the same time may appear on the network but are not connected. We construct the network with a threshold of 0.015. The reason for the lower threshold is that the correlation matrix is averaged across multiple stocks and periods. The community structure is detected using the Information-theoretic method (Infomap) on the first decile network. The modularity is 0.312 for this network. Depending on the community structure, the network nodes are colored. The red nodes correspond to the DIMs and the green nodes correspond to the DSMs and some FRMs. The blue nodes are mainly FRMs. The DIMs with large transaction sizes form a red cluster, and the remaining small DSMs are strongly forming other green clusters. Through the community structure of the network, we confirm the herding of member firms at the level of each member.

\section{Inventory variation and return}
Prior to the previous section, we only dealt with the relationship between the inventories of members. Thus, we will analyze the connection between the stock prices and the inventory variations in this section. The use of random matrix theory and a cross-sectional regression will reveal the connection between inventories and stock prices. Many researchers have studied a correlation matrix of stock returns using random Matrix Theory (RMT)\cite{Lillo2008,Shen2012,Aste2010,Zhou2012,Han2017,Zhao2018,Wang2018}. They have found that the largest eigenvalue has some information that cannot be explained by the hypothesis of a random matrix. The eigenvectors of the largest eigenvalues contain some information about the market trend or industrial sectors. There also have been attempts to apply RMT to the inventory variation of investors. W.X. Zhou et al.\cite{Zhou2012} studied the dynamics of the inventory variation of traders and found that the largest eigenvector is linearly related to returns on the Shenzhen Stock Exchange. F. Lillo et al.\cite{Lillo2008} analyzed the dynamics of the inventory variation of the member firms on the Madrid Stock Exchange. They showed that the factor, which is the projection of the inventory variation on the eigenvector of the largest eigenvalue, is linearly related to the price return. Like previous studies, we also analyze the correlation matrix of the inventory variation using RMT. To reduce the noise and to get meaningful information in the correlation matrix, we employ RMT \cite{Laloux1999} as follows:
\begin{eqnarray}
    \rho_c(\lambda) = \frac{1}{N}\frac{dn(\lambda)}{d\lambda}\\
    \rho_c(\lambda) = \frac{Q}{2\pi \sigma ^2} \frac{\sqrt{(\lambda_{max}-\lambda)(\lambda-\lambda_{min})}}{\lambda}\\
    \lambda_{min}^{max} = \sigma ^2(1+1/Q\pm 2\sqrt{1/Q})
\end{eqnarray}
where $n(\lambda)$ is the number of eigenvalues smaller than $\lambda$. $N$ is the total number of eigenvalues, which, in this case, is the number of member firms. Eq. 5 is the density of the eigenvalue and Eq. 6 is the density of the eigenvalue from the hypothesis of a random matrix, where $Q=T/N\simeq 247/62\simeq 3.98$. The average length of time series is 247 because we analyze the data yearly. Time series $x$ of the inventory variation has a dimension of $N\times T$, and so the correlation matrix of the inventory variation has a dimension of $N\times N$. In this study, $N$ is approximately 62, which is the number of member firms.

\begin{eqnarray}
    Factor(t) = \sum _i x_i(t) u_i(\lambda _1)(t)
\end{eqnarray}
$x_i$ is the inventory variation of stock $i$, $\lambda_1$ is the largest eigenvalue, and $u_i(\lambda_1)$ is the eigenvector of the largest eigenvalue.

We find the extent to which a factor describes the price return, depending on the size of the market capitalization of the stock. We verify that the factor is linearly related to the price return. Fig. 7 (a) shows the eigenvalue spectrum compared to the random matrix hypothesis. We confirm that there is the largest eigenvalue that is not described by a random matrix. In Fig. 7 (b), we find that the 1st decile that represents the largest market capitalization group (darker) has a larger correlation and eigenvalue than the smallest group (lighter). Fig. 7 (b) is the average values over the ten respective deciles. The larger the market capitalization is, the more the factor explains the return. This is because the absolute value of the trend ($T$) is large, as shown in Fig. 4 (b). Because of the large difference between the trends of each investor type, the differences in the characteristics between member companies become clear and eventually explain the return as a correlation matrix of the inventory variation.

\begin{figure*}[h!]
\includegraphics[scale=0.65]{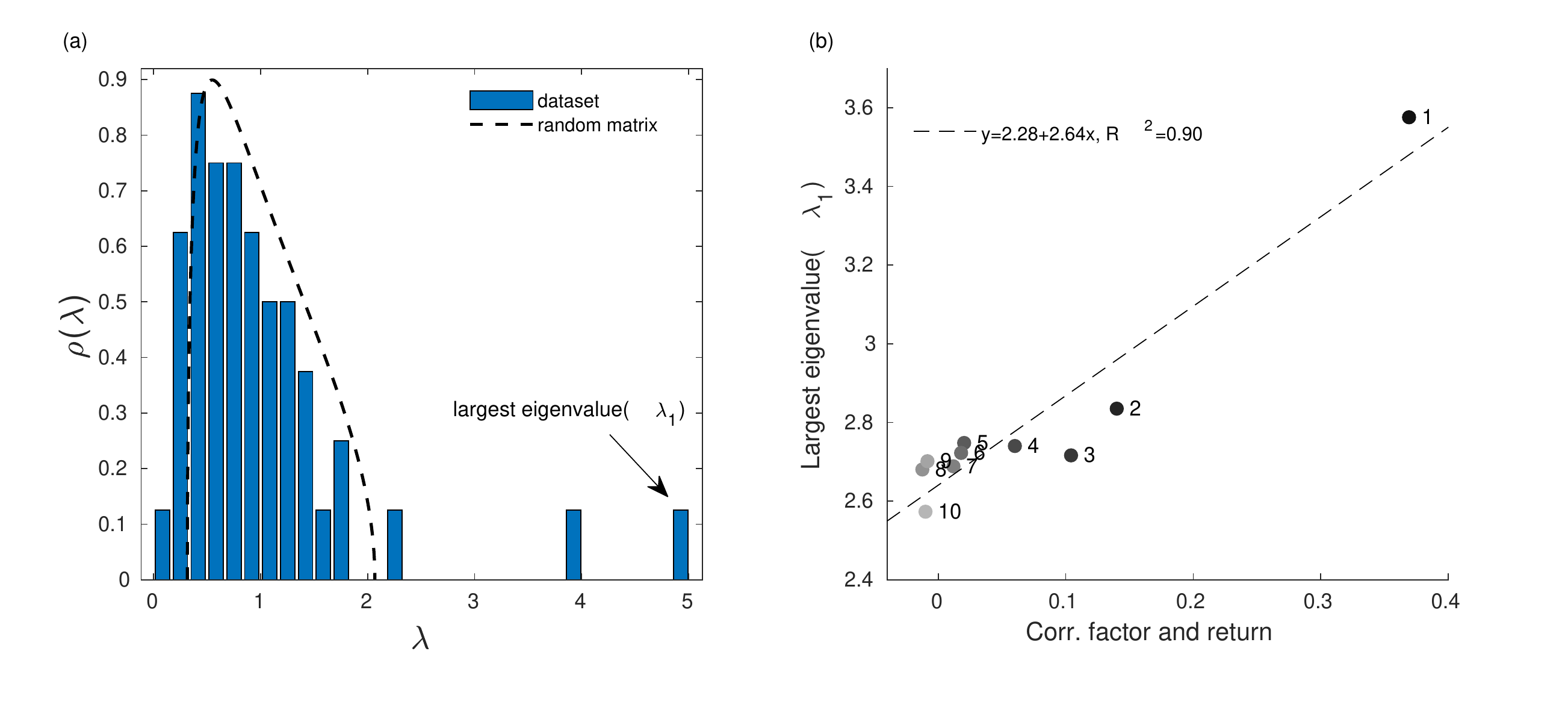}
\caption{Random matrix theory. (a) Distribution of the eigenvalues of the correlation matrix of inventory variation(histogram) and the distribution of the eigenvalues of the random matrix(dotted line). (b) Largest eigenvalue. and correlation between the factor \& return. The 1st-10th decile groups. The larger market capitalization group is, the darker the color is.}
\end{figure*}

Furthermore, we quantitatively analyze how the herding of members affects the stock price using a cross-sectional regression as follows: 

\begin{eqnarray}
    R_{it}-R_{ft}=\alpha+\beta_1 (R_{Mt}-R_{ft}) + \beta_2 H_{DSM} \\
    + \beta_3  H_{DIM} + \beta_4 H_{FRM} + Year
\end{eqnarray}

where $R_{it}$ is the return of stock $i$ at time $t$ and $R_{ft}$ is the return of a risk free asset at time $t$. We use the Korea 10-years bond yield rate as the risk-free asset. $H_{DIM}$ means the herding of DSMs. $H_{DIM} and H_{FRM}$ mean those of DIMs and FRMs, respectively. $Year$ is the dummy variable representing the yearly effects. As shown in Table 3, in addition to market factor, herding also has a significant relationship with the stock price. The DIMs have negative coefficients. On the other hand, the herding of DSMs and FRMs has positive coefficients. In the cross-sectional regression, we find that the R-squared value increases when considering the herding of members rather than just the market factor. We confirm that the inventory variation possesses information about the stock price and how much the herding of members affects the price through this section.

\begin{table}[]
\begin{center}
\caption{Table of the cross-sectional regression results. The independent variables are the market factor, the herding of domestic members, and the herding of foreign members. The dependent variable is the price return of each stock.}
\begin{tabular}{lclc}
\toprule
\textbf{Dep. Variable:}    &     returns      & \textbf{  R-squared:         } &      0.410   \\
\textbf{Model:}            &       OLS        & \textbf{  Adj. R-squared:    } &      0.410   \\
\textbf{Method:}           &  Least Squares   & \textbf{  F-statistic:       } &  1.632e+04   \\
\textbf{  Log-Likelihood:    } &  7.8233e+05  & \textbf{Df Model:}         &          14 \\
\textbf{No. Observations:} &      329362      & \textbf{  AIC:               } &  -1.565e+06  \\
\textbf{Df Residuals:}     &      329347      & \textbf{  BIC:               } &  -1.564e+06  \\
\bottomrule \hline
\end{tabular}
\begin{tabular}{lcccccc}
                         & \textbf{coef} & \textbf{std err} & \textbf{t} & \textbf{P$>$$|$t$|$} & \textbf{[0.025} & \textbf{0.975]}  \\
\midrule \hline
\textbf{$\alpha$}       &       0.0002  &        0.000     &     1.810  &         0.070        &    -1.95e-05    &        0.000     \\
\textbf{C(year)[T.2008]} &       0.0002  &        0.000     &     1.053  &         0.293        &       -0.000    &        0.001     \\
\textbf{C(year)[T.2009]} &      -0.0007  &        0.000     &    -3.761  &         0.000        &       -0.001    &       -0.000     \\
\textbf{C(year)[T.2010]} &      -0.0005  &        0.000     &    -2.812  &         0.005        &       -0.001    &       -0.000     \\
\textbf{C(year)[T.2011]} &      -0.0001  &        0.000     &    -0.588  &         0.556        &       -0.000    &        0.000     \\
\textbf{C(year)[T.2012]} &      -0.0007  &        0.000     &    -3.773  &         0.000        &       -0.001    &       -0.000     \\
\textbf{C(year)[T.2013]} &   -7.659e-05  &        0.000     &    -0.415  &         0.678        &       -0.000    &        0.000     \\
\textbf{C(year)[T.2014]} &       0.0001  &        0.000     &     0.680  &         0.496        &       -0.000    &        0.000     \\
\textbf{C(year)[T.2015]} &       0.0004  &        0.000     &     2.328  &         0.020        &     6.78e-05    &        0.001     \\
\textbf{C(year)[T.2016]} &      -0.0007  &        0.000     &    -3.936  &         0.000        &       -0.001    &       -0.000     \\
\textbf{C(year)[T.2017]} &      -0.0003  &        0.000     &    -1.711  &         0.087        &       -0.001    &     4.62e-05     \\
\textbf{Market}          &       0.9326  &        0.002     &   400.864  &         0.000        &        0.928    &        0.937     \\
\textbf{$H_{DSM}$}   &       0.0014  &        9e-05     &    15.231  &         0.000        &        0.001    &        0.002     \\
\textbf{$H_{DIM}$}   &      -0.0285  &        0.000     &  -225.522  &         0.000        &       -0.029    &       -0.028     \\
\textbf{$H_{FRM}$}         &       0.0019  &     8.47e-05     &    22.008  &         0.000        &        0.002    &        0.002     \\
\bottomrule
\hline
\end{tabular}
\end{center}
\end{table}

\section{Conclusion}
We analyze the trading characteristics of the member firms on the Korea Exchange. We deal with the difficulty of identifying the members by comparing the inventory variations of three types of investors. The properties of members are determined by their correlations with individuals, institutions, and foreigners. We also measure the directionality and trend to understand the dynamics of the member firms. The foreign members tend to trade in a one-way direction and trade in the same direction as a price movement. DIMs have a weak trading direction and trade in the opposite direction of a price movement. The herding of members moves in the opposite direction of a price change, unlike the common herding of investors. While FRMs do weak herding and move in the direction of a price movements. We construct a network of member firms from which we identify the connections between the members with similar trading characteristics. The community detection shows that the members are clustered into three groups. Random matrix theory reveals that the correlation matrix of inventory variation has a lot of information about the price dynamics, especially in the first decile. In addition, the cross-sectional regression shows that the herding of each group helps to price the stocks. This study will help investors understand the dynamics of domestic and foreign members because investors can access the daily inventory variations of members. This study can also help investors understand what strategies their members are using to trade. The dependency on the market capitalization of the factor related to the price return also means that other strategies are needed because the trading strategies of the member firms depend on the market capitalization of the stocks.

\section*{Acknowledgments}
This study was supported by research fund from Chosun University (K206026014-1, 2017)

\bibliographystyle{unsrt}  


\end{document}